\documentclass[11pt, a4paper, Physsubmission, Phys]{SciPost} 

\hypersetup{
    colorlinks,
    linkcolor={red!50!black},
    citecolor={blue!50!black},
    urlcolor={blue!80!black}
}

\usepackage[bitstream-charter]{mathdesign}
\usepackage{wrapfig}
\usepackage{enumitem}
\usepackage{xspace}

\usepackage{subfig}
\usepackage{amsmath}
\usepackage{gensymb}
\usepackage{mathrsfs}
\usepackage{lineno}

\DeclareSymbolFont{usualmathcal}{OMS}{cmsy}{m}{n}
\DeclareSymbolFontAlphabet{\mathcal}{usualmathcal}

\newcommand{\sib}[1]{{\sc Sibyll}~#1\xspace}
\newcommand{\qgsii}{{\sc Qgsjet~II-04}\xspace}
\newcommand{\eposlhc}{{\sc Epos-lhc}\xspace}

\newcommand{\Xmax}{\ensuremath{X_{\rm max}}\xspace}

\newcommand{\meanXmax}{\ensuremath{\langle X_{\rm max}\rangle}\xspace}
\newcommand{\erange}[2]{\ensuremath{10^{#1}-10^{#2}~{\rm eV}}\xspace}
\newcommand{\lgE}[1]{\ensuremath{10^{#1}~{\rm eV}}\xspace}

\newcommand{\twodimref}{\ensuremath{[X_\text{max}^\text{Ref},~S^\text{Ref}(1000)]}\xspace}

\begin{document}

\begin{center}{\Large \textbf{
Probing hadronic interaction models with the hybrid data of the Pierre Auger Observatory\\
}}\end{center}

\begin{center}
Jakub Vícha\textsuperscript{1$\star$} for the Pierre Auger Collaboration\textsuperscript{2$\dagger$}
\end{center}

\begin{center}
{\bf 1} Institute of Physics of the Czech Academy of Sciences
\\
{\bf 2} Observatorio Pierre Auger, Av. San Martín Norte 304, 5613 Malargüe, Argentina
\\

$\star$ vicha@fzu.cz\\
$\dagger$spokespersons@auger.org, \href{https://www.auger.org/archive/authors\_2022\_05.html}{https://www.auger.org/archive/authors\_2022\_05.html}
\end{center}

\begin{center}
\today
\end{center}


\definecolor{palegray}{gray}{0.95}
\begin{center}
\colorbox{palegray}{
  \begin{tabular}{rr}
  \begin{minipage}{0.1\textwidth}
    \includegraphics[width=30mm]{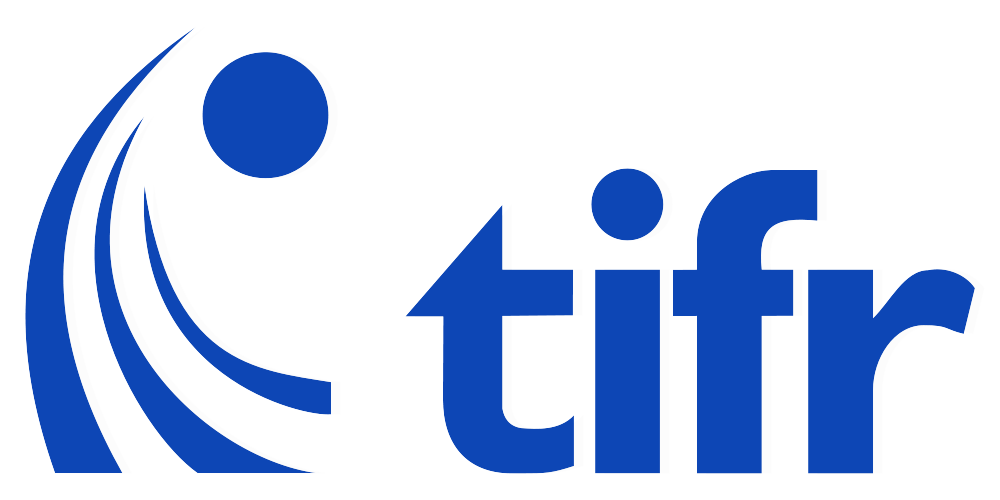}
  \end{minipage}
  &
  \begin{minipage}{0.85\textwidth}
    \begin{center}
    {\it 21st International Symposium on Very High Energy Cosmic Ray Interactions (ISVHE- CRI 2022)}\\
    {\it Online, 23-28 May 2022} \\
    \doi{10.21468/SciPostPhysProc.?}\\
    \end{center}
  \end{minipage}
\end{tabular}
}
\end{center}

\section*{Abstract}
{\bf
\bf
\boldmath
Presently large systematic uncertainties remain in the description of hadronic interactions at ultra-high energies and a fully consistent description of air-shower experimental data is yet to be reached. 
The amount of data collected by the Pierre Auger Observatory using simultaneously the fluorescence and surface detectors in the energy range $\mathbf{10^{18.5}-10^{19.0}}$~eV has provided opportunity to perform a multi-parameter test of model predictions.
We apply a global method to simultaneously fit the mass composition of cosmic rays and adjustments to the simulated depth of shower maximum, and hadronic signal at ground level. 
The best description of the hybrid data is obtained for a deeper scale of simulated depth of shower maximum than predicted by hadronic interaction models tuned to the LHC data. 
Consequently, the deficit of the simulated hadronic signal at ground level, dominated by muons, is alleviated with respect to the unmodified hadronic interaction models.
Because of the size of the adjustments to simulated depth of shower maximum and hadronic signal and the large number of events in the sample, the statistical significance of these assumed adjustments is large, greater than 5$\mathbf{\sigma_\text{stat}}$, even for the combination of the systematic experimental shifts within 1$\mathbf{\sigma_\text{sys}}$ that are the most favorable for the models.
}

\vspace{10pt}
\noindent\rule{\textwidth}{1pt}
\tableofcontents\thispagestyle{fancy}
\noindent\rule{\textwidth}{1pt}
\vspace{10pt}

\vspace{-1cm}
\section{Problems of Models of Hadronic Interaction}
\label{sec:intro}
\begin{figure}[ht] 
\centering
\def\h{0.42}
\subfloat{\includegraphics[height=\h\textwidth]{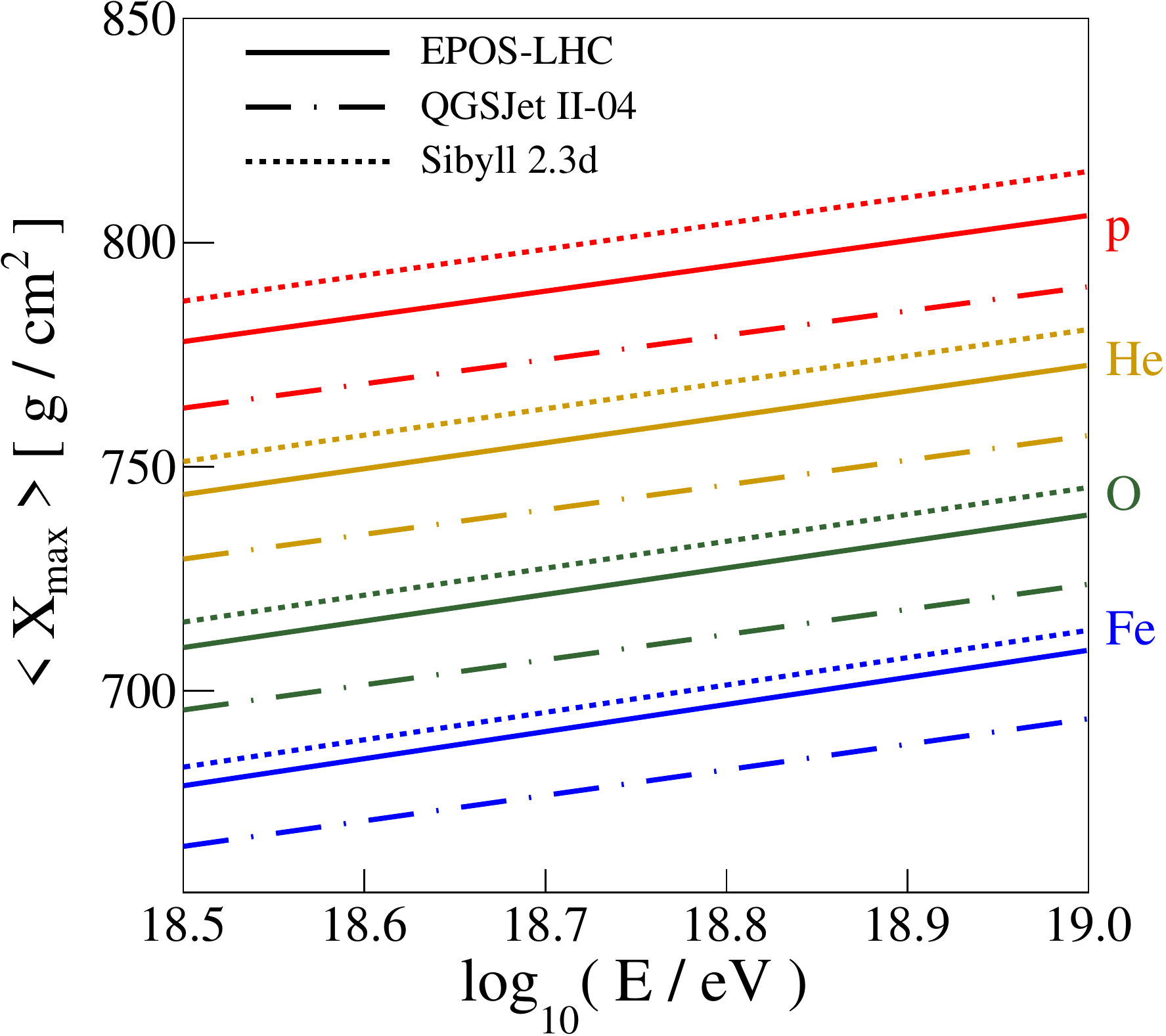}}
\hspace{0.3cm}
\subfloat{\includegraphics[height=\h\textwidth]{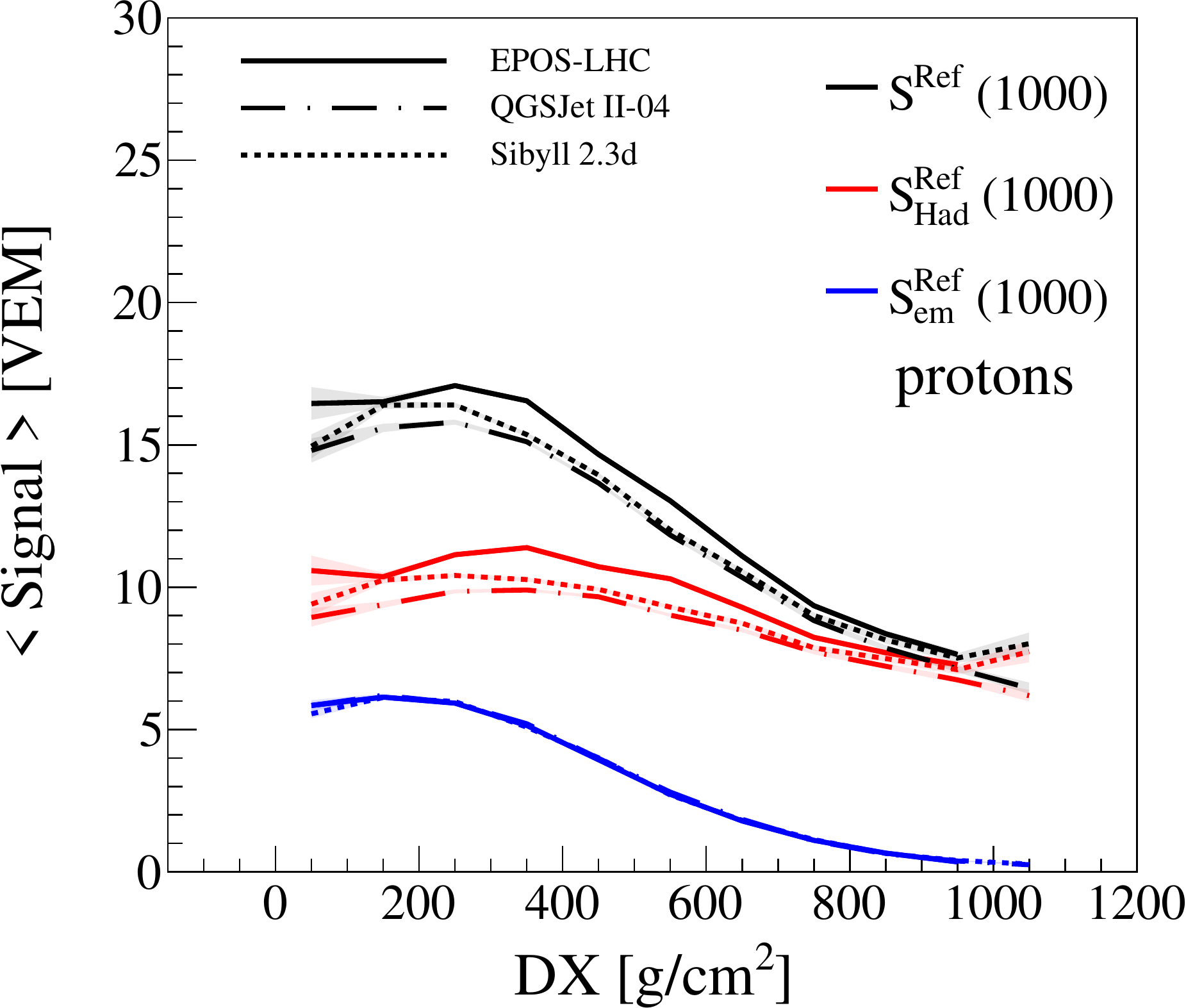}}
\caption{Left: the energy evolution of the mean of the $X_{\rm max}$ distribution predicted for three models of hadronic interactions and four primary species. Right: the dependence of average signal at the Pierre Auger Observatory induced in SD at 1000\,m from the shower core (black) on the distance of \Xmax to the ground ($DX=880~$g/cm$^{2}/\cos(\theta)-X_{\rm max}$) for proton showers of energy $10^{18.5}-10^{19.0}$~$\mathrm{eV}$. The hadronic ($S^\text{Ref}_\text{Had}$) and electromagnetic ($S^\text{Ref}_\text{em}$) parts of the signal are shown in red and blue, respectively. The signals are corrected to the reference energy $10^{18.7}$~$\mathrm{eV}$. From \cite{MethodAugerDataICRC21}.}
\vspace{-0.5cm}
\label{fig:MCdifferences}
\end{figure}

The current models of hadronic interactions (HI models) are known to have problems consistently describing both the ground signals and the longitudinal shower development \cite{TestingHadronicInteractions,InclinedMuons,AmigaMuons} using the hybrid data of the surface detectors (SD) and fluorescence detectors (FD) at the Pierre Auger Observatory \cite{PACosmicObservatory}.
The ground signal predicted by Monte Carlo (MC) simulation is found lower than measured, with an indication of increasing with energy when the results of more experiments are combined \cite{WHISParticle}.
All these tests of HI models are based on the assumption that the scale of depth of shower maximum (\Xmax) predicted by a given HI model is correct.
However, at the energy $10^{18.7}$~eV, the uncertainties on the predicted \Xmax scale, $\langle X_\text{max} \rangle$, are larger than about one third of the difference between the two extremes - protons and iron nuclei, see the differences between the three HI models, \eposlhc \cite{EposLHC}, \qgsii \cite{Qgsjet}, \sib{2.3d} \cite{Sibyll} on the left panel of Fig.~\ref{fig:MCdifferences}, see also \cite{XmaxUncertainty}.

\section{Extended Testing of Hadronic Interaction Models}
In this work, we consider a more complex test of the HI models in the energy range $10^{18.5-19.0}$~$\mathrm{eV}$.
This test is motivated by the differences observed in the predictions obtained with HI models for $\langle X_\text{max}\rangle$ and hadronic part $S_\text{Had}$ of the total ground signal $S(1000)$ at 1000~m from the shower core at the Pierre Auger Observatory, see Fig.~\ref{fig:MCdifferences}.
These differences are approximately primary and energy independent.
The differences in the hadronic signals are $DX$ (zenith-angle) dependent.
These features suggest to leave both \Xmax and $S_\text{Had}(\theta)$ free in the fitting procedure. 

\begin{figure*}
  \centering
  \def\h{0.22}
  \includegraphics[width=\textwidth]{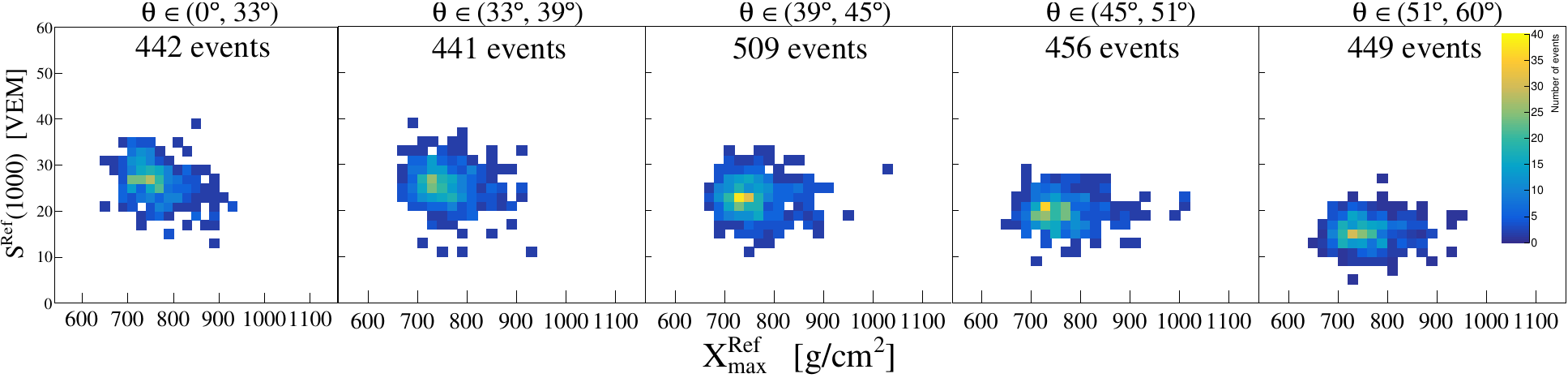}
  \caption{Two-dimensional distributions of $S(1000)^{\rm Ref}$ and $X_{\rm max}^{\rm Ref}$ measured by the Pierre Auger Observatory in the energy range \erange{18.5}{19.0} in five zenith-angle ($\theta$) intervals. From \cite{MethodAugerDataICRC21}.}
\vspace{-0.2cm}
  \label{DataPlots}
\end{figure*}

We fit simultaneously five two-dimensional distributions of measured $S(1000)$ and \Xmax, see Fig.~\ref{DataPlots}, that are corrected for their energy evolution to a reference energy $10^{18.7}$~$\mathrm{eV}$, see \cite{MethodAugerDataICRC21} for more details.
The influence of changing \Xmax scale on $S(1000)$ is incorporated in the method.
Implicitly, the nearly model-independent information on mass composition from the correlation between $S(1000)$ and \Xmax, see \cite{MixedAnkle}, is naturally incorporated in the method.


\section{Results of the Test}
\label{sec:results}
\begin{figure}[ht!]
\vspace{-0.5cm}
  \centering
  \def\h{0.35}
\subfloat[]{\includegraphics[height=\h\textwidth]{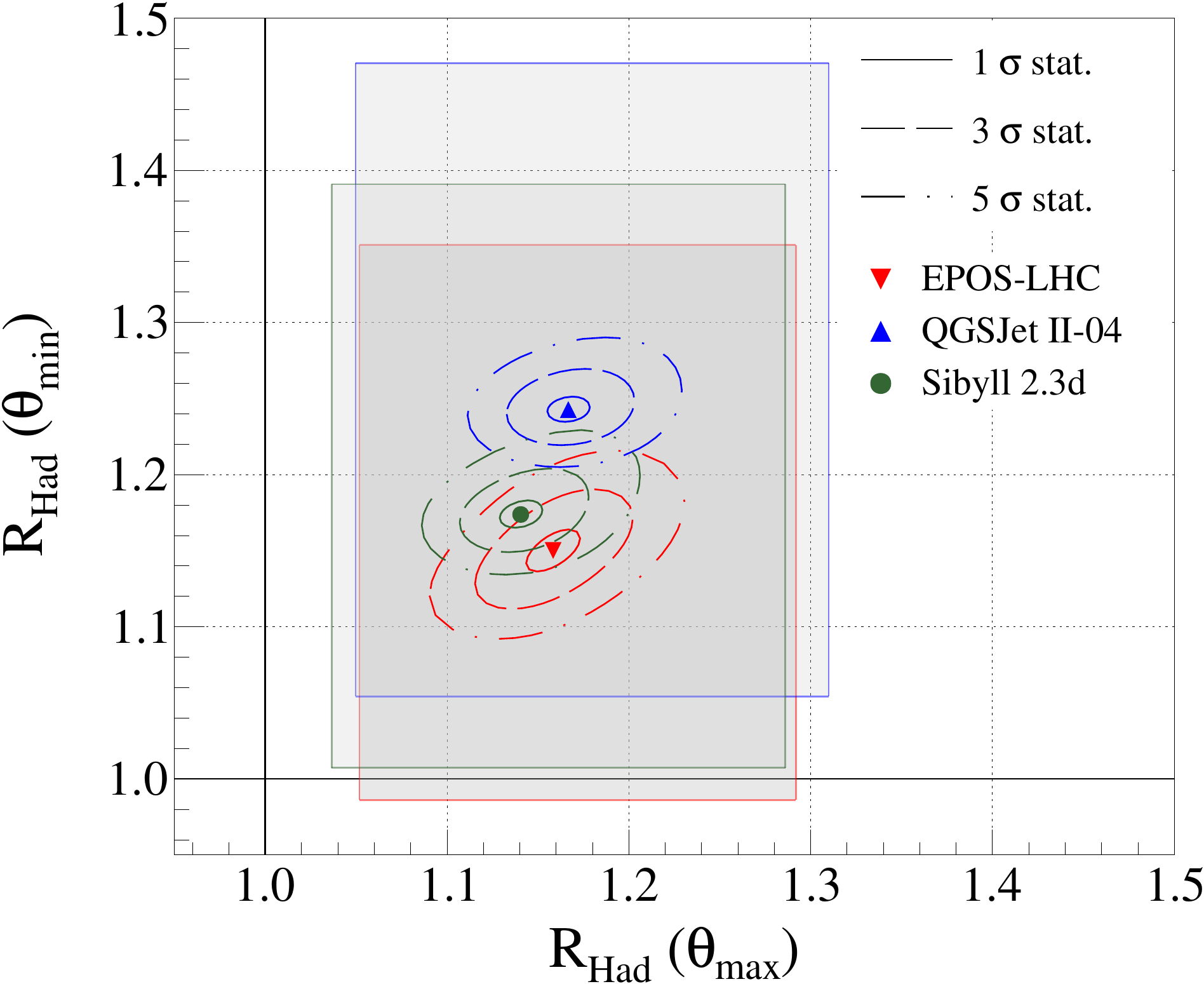}}
\hspace{0.5cm}
\subfloat[]{\includegraphics[height=\h\textwidth]{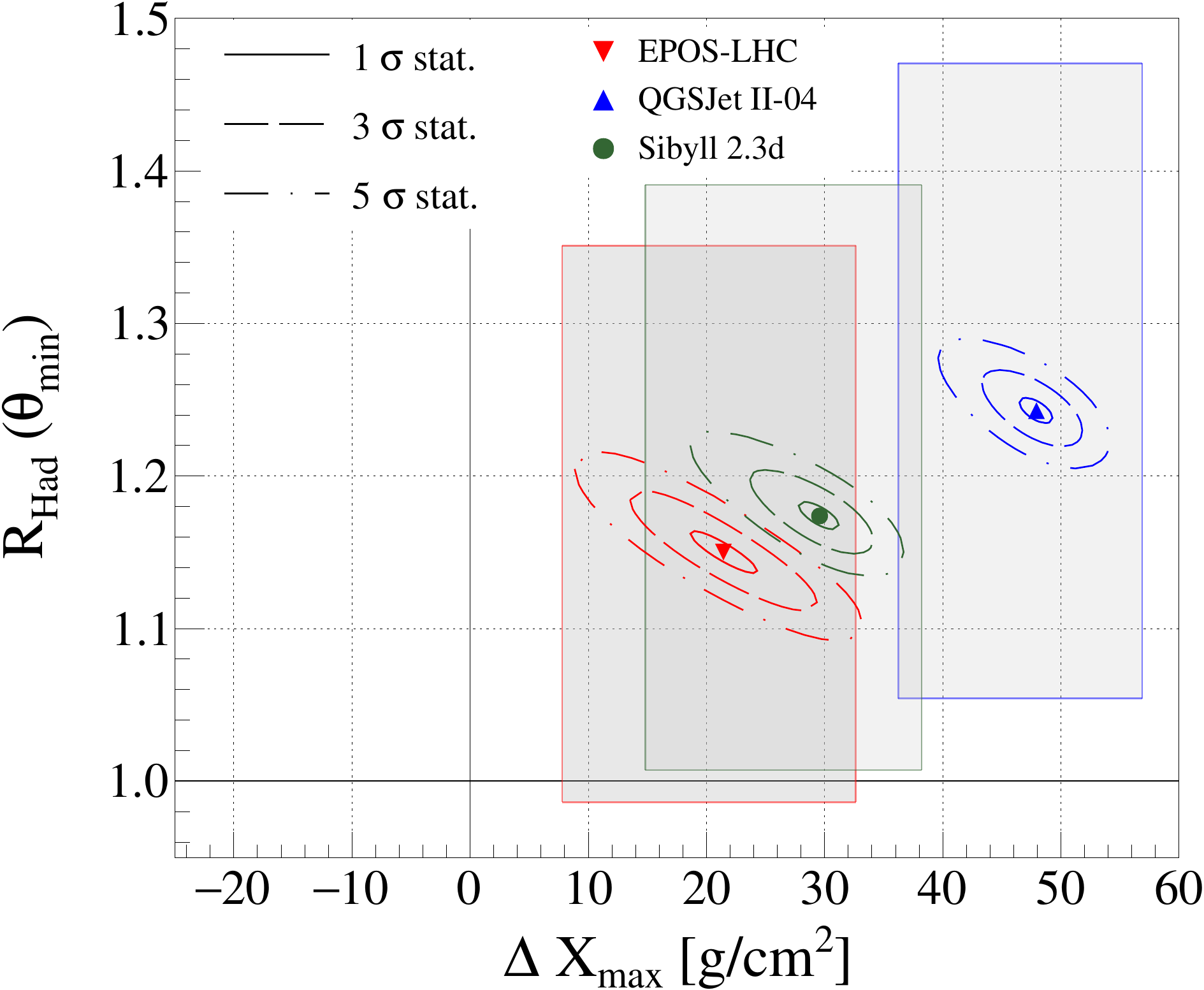}}\\
\vspace{-0.3cm}
\subfloat[]{\includegraphics[height=\h\textwidth]{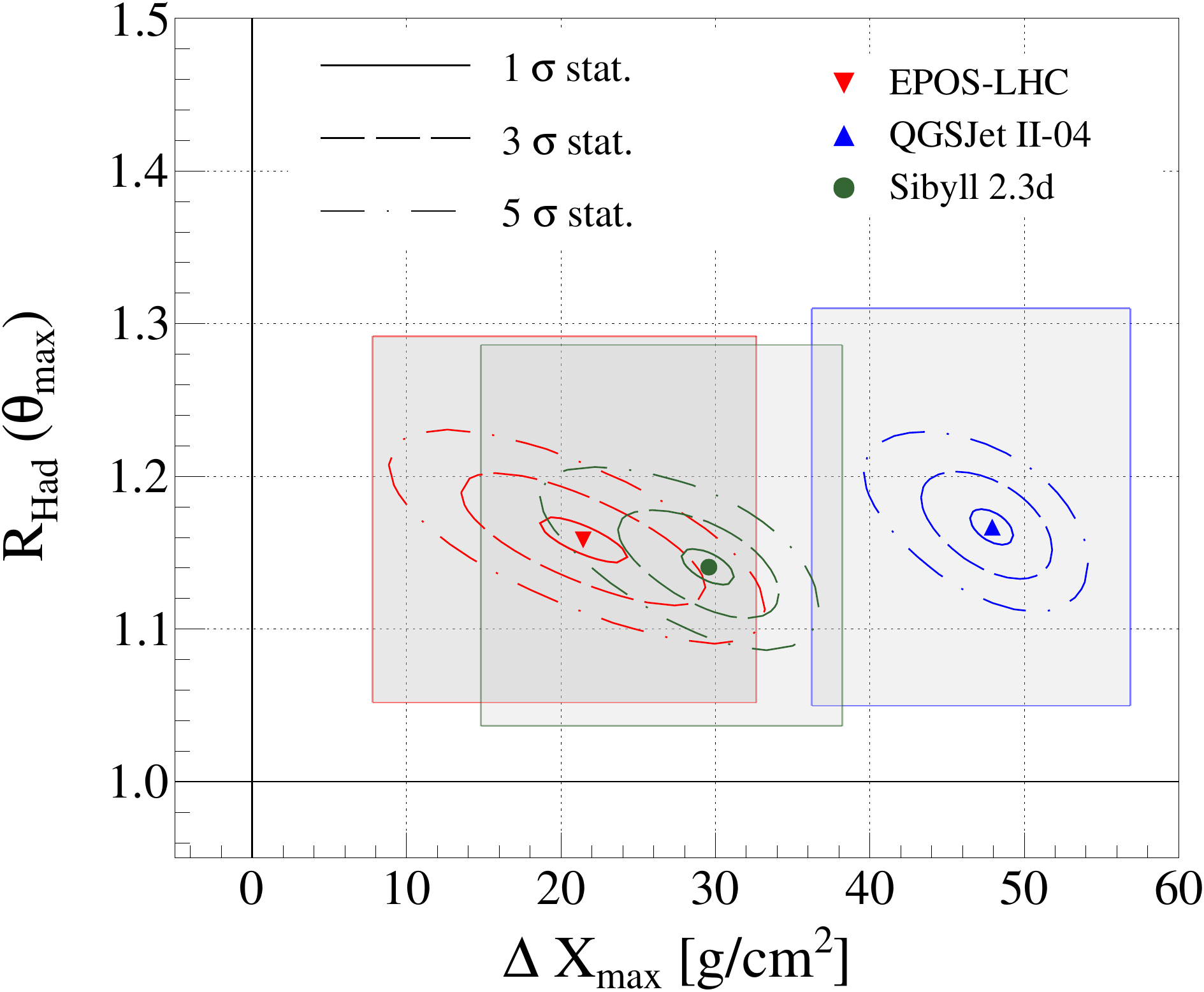}}
\hspace{0.5cm}
\subfloat[]{\includegraphics[height=\h\textwidth]{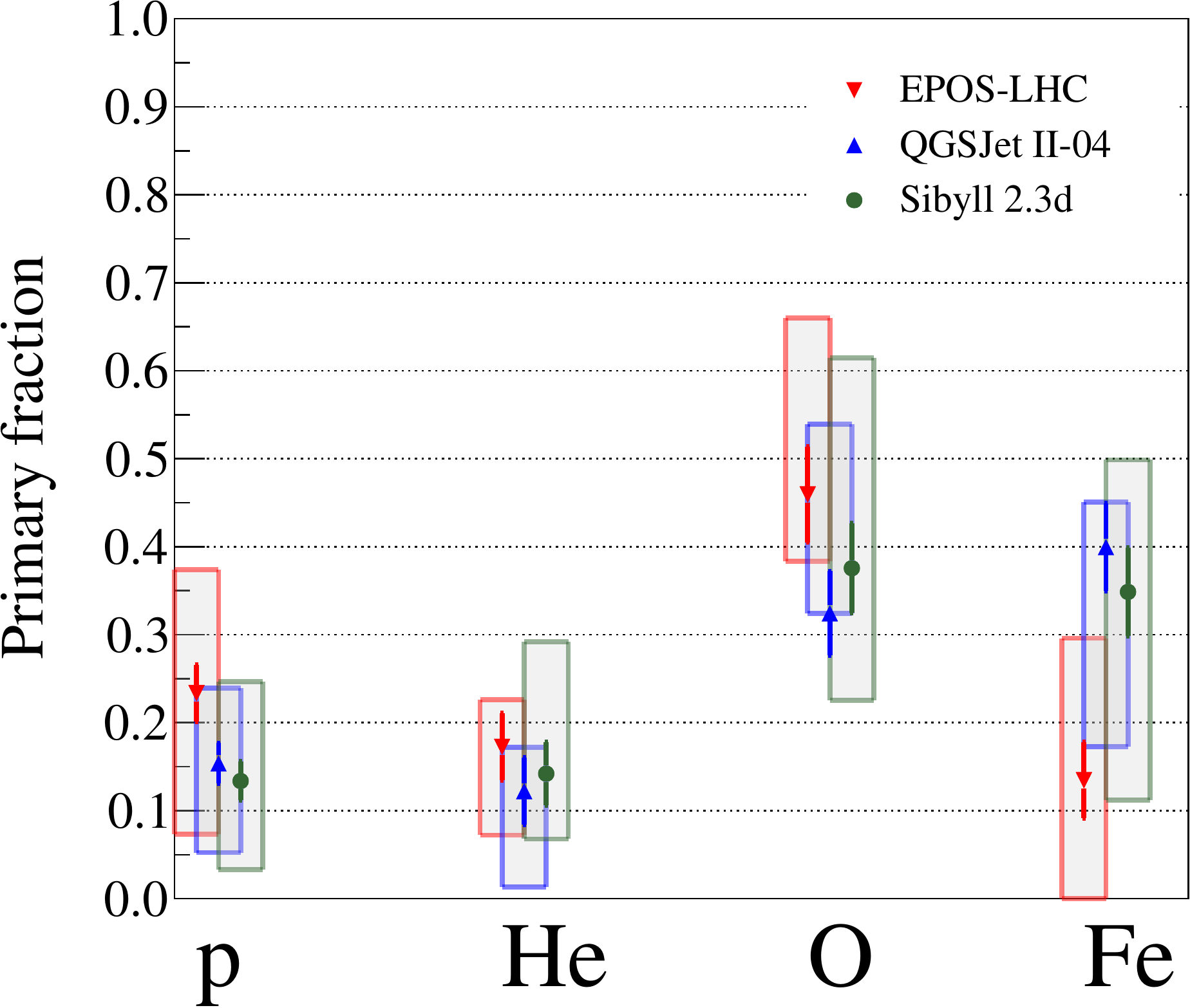}}
\caption{Results of the fits to \twodimref distributions with MC templates for \eposlhc, \qgsii and \sib{2.3d}. The relations between the fit parameters $R_\text{Had}(\theta_\text{min})$, $R_\text{Had}(\theta_\text{max})$ and $\Delta X_\text{max}$ are shown in (a), (b), (c) panels and the fitted primary fractions in panel (d). The contours denote regions with 1$\sigma$, 3$\sigma$ and 5$\sigma$ statistical uncertainties. The gray bands indicate the size of the total systematic uncertainties. From \cite{MethodAugerDataICRC21}.}
  \label{Fig:FitParameters}
\vspace{-0.2cm}
\end{figure}

The best description of the hybrid data of the Pierre Auger Observatory is obtained modifying the MC templates by shifting \Xmax by a parameter $\Delta X_\text{max}$ and rescaling the hadronic component at two extreme zenith angles by factors $R_\text{Had}(\theta_\text{min})$ and $R_\text{Had}(\theta_\text{max})$ as shown in Fig.~\ref{Fig:FitParameters} with statistical and systematic errors.
The overall description of the five two-dimensional distributions by the fits is achieved with $p$-value~$\simeq$~2.6\% for \eposlhc, $\simeq$~3.6\% for \qgsii, and $\simeq$~18.0\% for \sib{2.3d} based on detailed MC-MC tests.

\begin{figure}[ht!]
\vspace{-0.5cm}
  \centering
    \def\h{0.44}
\subfloat{\includegraphics[height=\h\textwidth]{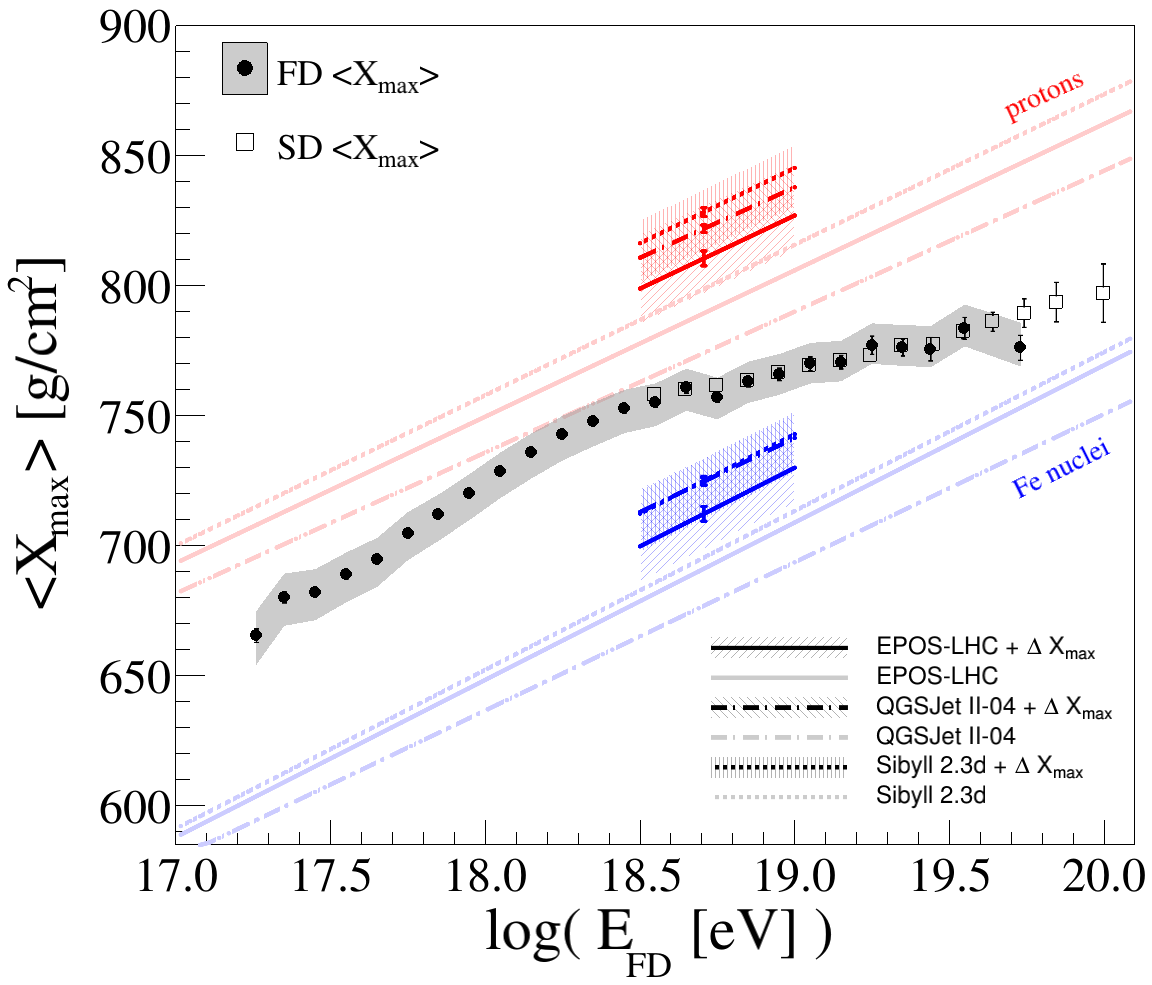}}
\hspace{0.2cm}
\subfloat{\includegraphics[height=\h\textwidth]{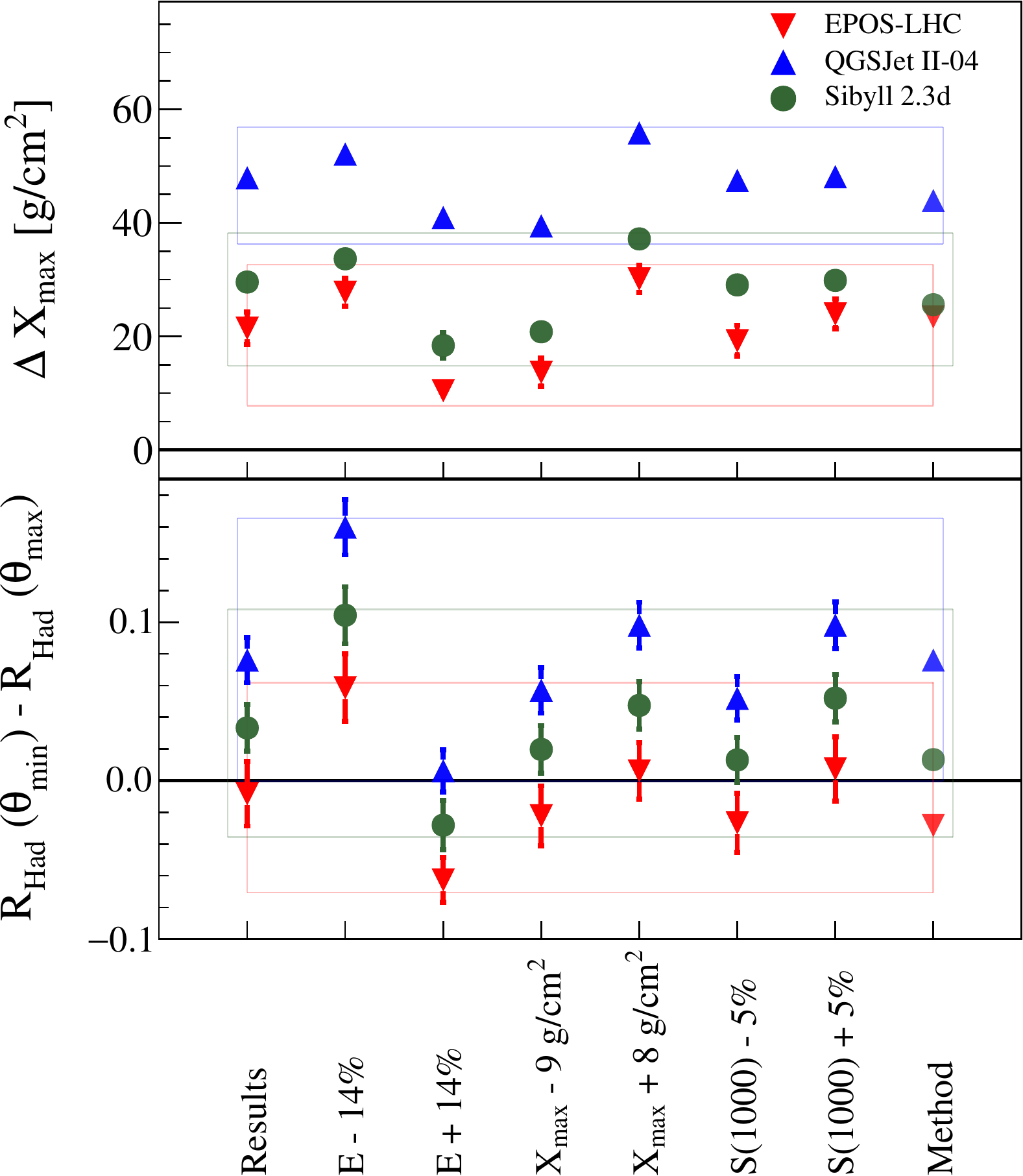}}
\caption{Left: the energy evolution of the mean \Xmax measured at the Pierre Auger Observatory (black) using FD \cite{XmaxICRC19} and SD \cite{DeltaICRC19}. 
The adjusted MC predictions on \meanXmax obtained in this work to improve the description of hybrid data (SD and FD) are shown in red and blue for protons and iron nuclei, respectively. The bands correspond to the systematic uncertainties of $\Delta$\Xmax. The lighter color lines indicate the unmodified MC predictions. Right: the best fit results on the $\Delta$\Xmax and the adjustment to the $S_\text{Had}$ attenuation for the individual systematic effects. The bands illustrate the total systematic uncertainty summed in quadrature. From \cite{MethodAugerDataICRC21}.}
  \label{Fig:XmaxMoments}
\end{figure}

The optimum description of the data is achieved for a deeper \Xmax scale than predicted by HI models, see also left panel of Fig.~\ref{Fig:XmaxMoments}. Consequently, heavier mass composition and smaller rescaling of the hadronic component is obtained than using previous tests \cite{TestingHadronicInteractions} and mass composition fits \cite{XmaxIcrc2017}.

\paragraph{Systematic Uncertainties}
\label{sec:systematics}
We have identified four dominant sources of systematic uncertainties influencing our results.
Three are related to the experimental uncertainties on the measurement of energy ($\pm14\%$), of \Xmax ($^{+8}_{-9}$~g/cm$^{2}$) and of $S(1000)$ ($\pm5\%$).
We consider also the uncertainty stemming from the MC-MC tests of the method. 
The total uncertainty is summed in quadrature and is plotted in Fig.~\ref{Fig:FitParameters} with shaded boxes.
The individual systematic contributions to $\Delta \Xmax$ and differences in the two fitted $R_\text{Had}$ are shown in the right panel of Fig.~\ref{Fig:XmaxMoments}.
The size of systematic uncertainty on the energy scale does not allow us to draw any conclusion about the muon spectrum from the $R_\text{Had}$ differences.  

\paragraph{Significance of Data/MC Inconsistency}
\label{sec:systematics}
We perform a scan in all possible combinations of the $1\sigma$ experimental systematic uncertainties and calculate the statistical significance as 

  \begin{equation}
    \sigma_{\rm STAT}=\sqrt{ \left(\frac{R_{\rm Had}(\theta_\text{min})-1}{\sigma_{\rm STAT}(R_{\rm Had}(\theta_\text{min}))}\right)^{2} + \left(\frac{R_{\rm Had}(\theta_\text{max})-1}{\sigma_{\rm STAT}(R_{\rm Had}(\theta_\text{max})}\right)^{2} + \left(\frac{\Delta X_{\rm max}}{\sigma_{\rm STAT}(\Delta X_{\rm max})}\right)^{2} }~,
    \label{EqStatSignificance}
  \end{equation}
  where $\sigma_{\rm STAT}(R_{\rm Had}(\theta_\text{min}))$, $\sigma_{\rm STAT}(R_{\rm Had}(\theta_\text{max}))$ and $\sigma_{\rm STAT}(\Delta X_{\rm max})$ are the statistical uncertainties of the fit for $R_{\rm Had}(\theta_\text{min})$, $R_{\rm Had}(\theta_\text{max})$ and $\Delta X_{\rm max}$, respectively.
  
\begin{figure}[ht!]
  \centering
    \def\h{0.36}
\subfloat{\includegraphics[height=\h\textwidth]{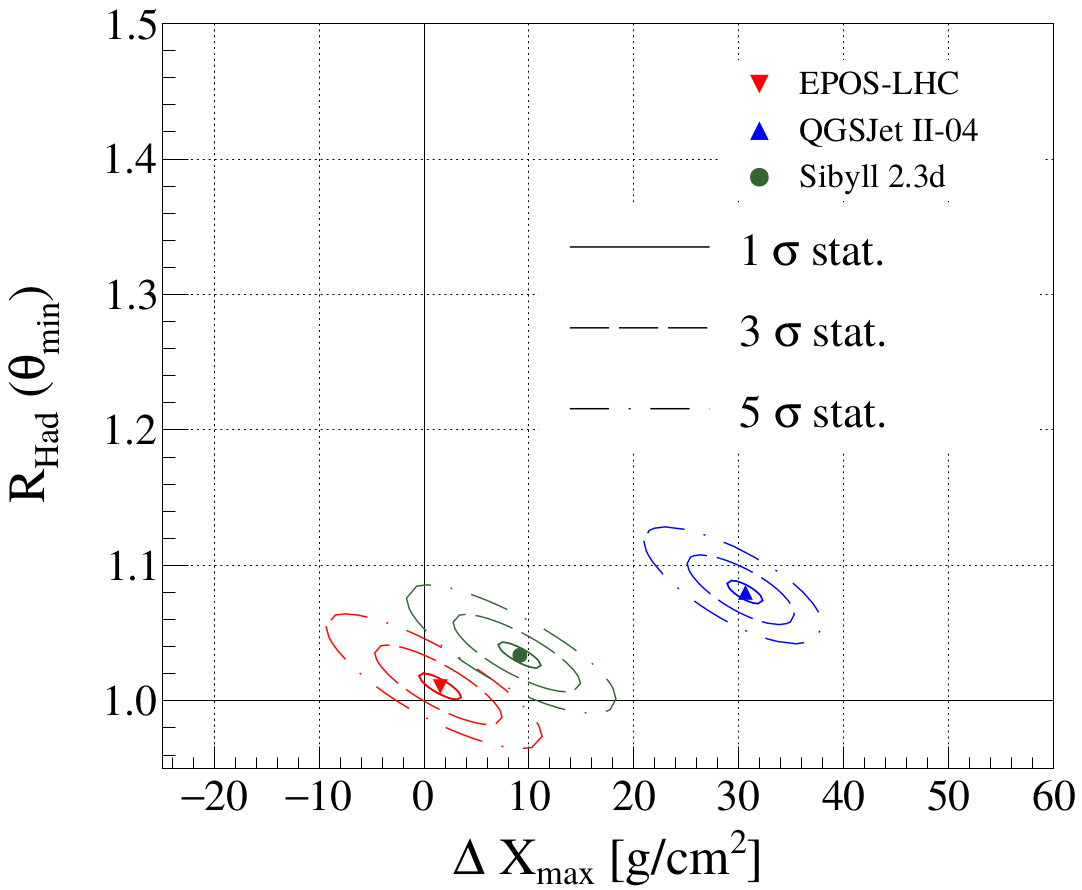}}
\hspace{0.2cm}
\subfloat{\includegraphics[height=\h\textwidth]{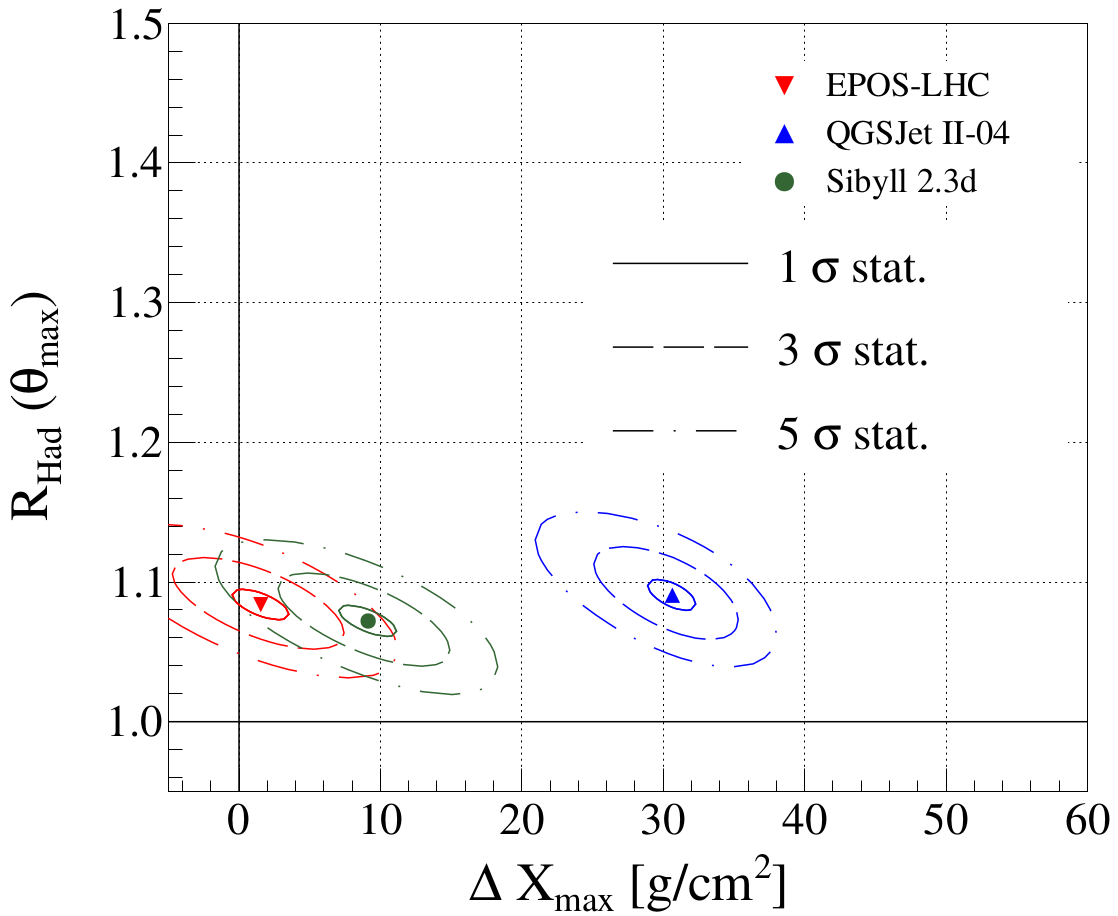}}
\caption{The MC adjustments obtained by fits to the measured data that are systematically corrected for energy +14\% and \Xmax - 9~g/cm$^{2}$ to minimize the total statistical significance from Eq.~(\ref{EqStatSignificance}). The contours denote regions with 1$\sigma$, 3$\sigma$ and 5$\sigma$ statistical uncertainties.}
\vspace{-0.2cm}
  \label{Fig:SystematicShifts}
\end{figure}

Even for the most favourable combination of the systematic uncertainties that minimizes $\sigma_\text{STAT}$ the significance value is still above 5$\sigma$.
This example is shown in Fig.~\ref{Fig:SystematicShifts}.

\section{Conclusions}
  The combined measurements of cosmic-rays with energies between \lgE{18.5} and \lgE{19.0} using the SD and FD of the Pierre Auger Observatory provide event statistics large enough to apply a novel method to test HI models.
  This method introduces three adjustments to the air-shower observables predicted by the HI models and simultaneously fits the fractions of four primary particles.  
  The results of the method are in tension with the HI models \eposlhc, \sib{2.3d} and \qgsii by more than 5$\sigma$ with much higher significance in case of the \qgsii.  
  The best description of the data is obtained when the $X_{\rm max}$ scale of the MC predictions is deeper by about 20 g/cm$^{2}$, 30 g/cm$^{2}$ and 50 g/cm$^{2}$ for \eposlhc, \sib{2.3d} and \qgsii, respectively.  
  Consequently, the differences between the HI models in the fitted mass composition fractions are decreased and the mass composition is heavier than obtained using the unmodified predictions of the three HI models.  
  For such heavier mass composition, the "muon problem" of the HI models is alleviated with respect to the previous studies, observing 15-25\% deficit of the hadronic component of simulated ground signal for all the three
  studied HI models.

\vspace{-0.2cm}

\section*{Acknowledgements}
This work is supported by the Czech Republic grants of MEYS CR: $\rm LTT18004$, $\rm LM2015038$, $\rm LM2018102$, $\rm CZ.02.1.01/0.0/0.0/16\_013/0001402$, $\rm CZ.02.1.01/0.0/0.0/18\_046/0016010$.

\bibliography{bibtex}

\begin{thebibliography}{10}
\providecommand{\url}[1]{\texttt{#1}}
\providecommand{\urlprefix}{URL }
\expandafter\ifx\csname urlstyle\endcsname\relax
  \providecommand{\doi}[1]{doi:\discretionary{}{}{}#1}\else
  \providecommand{\doi}{doi:\discretionary{}{}{}\begingroup
  \urlstyle{rm}\Url}\fi
\providecommand{\eprint}[2][]{\url{#2}}

\bibitem{TestingHadronicInteractions}
A.~Aab \emph{et~al.},
\newblock \emph{{Testing Hadronic Interactions at Ultrahigh Energies with Air
  Showers Measured by the Pierre Auger Observatory}},
\newblock Phys. Rev. Lett. \textbf{117}, 192001 (2016),
\newblock \doi{10.1103/PhysRevLett.117.192001}.

\bibitem{InclinedMuons}
A.~Aab \emph{et~al.},
\newblock \emph{Muons in air showers at the {Pierre} {Auger} {Observatory}:
  {Mean} number in highly inclined events},
\newblock Phys. Rev. D \textbf{91}, 032003 (2015),
\newblock \doi{10.1103/PhysRevD.91.032003}.

\bibitem{AmigaMuons}
A.~Aab \emph{et~al.},
\newblock \emph{{Direct measurement of the muonic content of extensive air
  showers between $2\cdot10^{17}$ and $2\cdot10^{18}$ eV at the Pierre Auger
  Observatory}},
\newblock Eur. Phys. J. C \textbf{210}, 751 (2020),
\newblock \doi{10.1140/epjc/s10052-020-8055-y}.

\bibitem{PACosmicObservatory}
{A. Aab et al.},
\newblock \emph{{The Pierre Auger Cosmic Ray Observatory}},
\newblock {Nucl. Instrum. Methods Phys. Res. A} \textbf{798}, 172  (2015),
\newblock \doi{doi.org/10.1016/j.nima.2015.06.058}.

\bibitem{WHISParticle}
J.~Albrecht \emph{et~al.},
\newblock \emph{{The Muon Puzzle in cosmic-ray induced air showers and its
  connection to the Large Hadron Collider}},
\newblock Astrophysics and Space Science \textbf{367}, 27 (2022),
\newblock \doi{10.1007/s10509-022-04054-5}.

\bibitem{EposLHC}
T.~Pierog \emph{et~al.},
\newblock \emph{{EPOS LHC}: Test of collective hadronization with data measured
  at the cern large hadron collider},
\newblock Phys. Rev. C \textbf{92}, 034906 (2015),
\newblock \doi{10.1103/PhysRevC.92.034906}.

\bibitem{Qgsjet}
S.~Ostapchenko,
\newblock \emph{{Monte Carlo treatment of hadronic interactions in enhanced
  Pomeron scheme: {QGSJET-II} model}},
\newblock Phys. Rev. D \textbf{83}, 014018 (2011),
\newblock \doi{10.1103/PhysRevD.83.014018}.

\bibitem{Sibyll}
F.~Riehn \emph{et~al.},
\newblock \emph{Hadronic interaction model sibyll 2.3d and extensive air
  showers},
\newblock Phys. Rev. D \textbf{102}, 063002 (2020),
\newblock \doi{10.1103/PhysRevD.102.063002}.

\bibitem{XmaxUncertainty}
R.~{Abbasi} and G.~{Thomson},
\newblock \emph{{$\langle X_{max}\rangle$ Uncertainty from Extrapolation of
  Cosmic Ray Air Shower Parameters}},
\newblock In \emph{Proceedings, Ultra-High Energy Cosmic Rays 2016}, p. 011015,
\newblock \doi{10.7566/JPSCP.19.011015} (2018).

\bibitem{MethodAugerDataICRC21}
{J. V\'icha for the Pierre Auger Collaboration},
\newblock \emph{{Adjustments to Model Predictions of Depth of Shower Maximum
  and Signals at Ground Level using Hybrid Events of the Pierre Auger
  Observatory}},
\newblock In \emph{{Proc., 37th International Cosmic Ray Conference,
  PoS(ICRC2021)310}},
\newblock \doi{10.22323/1.395.0310} (2021).

\bibitem{MixedAnkle}
A.~Aab \emph{et~al.},
\newblock \emph{Evidence for a mixed mass composition at the ‘ankle’ in the
  cosmic-ray spectrum},
\newblock Physics Letters B \textbf{762}, 288  (2016),
\newblock \doi{10.1016/j.physletb.2016.09.039}.

\bibitem{XmaxIcrc2017}
{J. Bellido for the Pierre Auger Collaboration},
\newblock \emph{{Depth of maximum of air-shower profiles at the Pierre Auger
  Observatory: Measurements above 10$^{17.2}$~{eV} and Composition
  Implications}},
\newblock In \emph{{Proc., 35th International Cosmic Ray Conference,
  PoS(ICRC2017)506}},
\newblock \doi{10.22323/1.301.0506} (2018).

\bibitem{XmaxICRC19}
{A. Yushkov for the Pierre Auger Collaboration},
\newblock \emph{{Mass composition of cosmic rays with energies above
  $10^{17.2}$ eV from the hybrid data of the Pierre Auger Observatory}},
\newblock In \emph{{Proc., 36th International Cosmic Ray Conference,
  PoS(ICRC19)482}},
\newblock \doi{10.22323/1.358.0482} (2019).

\bibitem{DeltaICRC19}
{C. J. T. Peixoto for the Pierre Auger Collaboration},
\newblock \emph{{Estimating the Depth of Shower Maximum using the Surface
  Detectors of the Pierre Auger Observatory}},
\newblock In \emph{{Proc., 36th International Cosmic Ray Conference,
  PoS(ICRC19)440}},
\newblock \doi{10.22323/1.358.0440} (2019).

\end{thebibliography}
\vspace{-0.2cm}



\nolinenumbers

\end{document}